\documentclass[preprint,showpacs,aps,amssymb,floatfix,prd,amsmath,preprintnumbers,showkeys]{revtex4} 
\usepackage{epstopdf}
\usepackage{capt-of}
\usepackage{graphicx}  % Include figure files
\usepackage{dcolumn}   % Align table columns on decimal point
\usepackage{bm}% bold math
\begin{document}
\input epsf.tex
%%%%%%%%%%%%%%%%%%%%%%%%%%%%%

\title{Tryon's conjecture and Energy and momentum of  Bianchi Type Universes\footnote{\scriptsize A part of this collaborative research  work was done during a  workshop conducted by IGIT Physics forum on  General Relativity, Gravitation and Cosmology  held at Indira Gandhi Intitute of Technology, Sarang, Odisha (India).}}

\author	{
	Prajyot Kumar Mishra\footnote{\scriptsize {Department of Electrical Engineering, Indira Gandhi Institute of Technology, Sarang, Dhenkanal, Odisha-759146, India, Email: kumarmishrap690@gmail.com}},
	Bibhudatta Panda\footnote{\scriptsize {Department of Electrical Engineering, Indira Gandhi Institute of Technology, Sarang, Dhenkanal, Odisha-759146, India, Email: aaryaphys@gmail.com}},
	Pradosh Ranjan Pattanayak\footnote{\scriptsize{Department of Computer Sciene Engineering and Applications, Indira Gandhi Institute of Technology, Sarang, Dhenkanal, Odisha-759146, India, Email:prpigitcse@gmail.com}}
	and
	Sunil Kumar Tripathy\footnote{\scriptsize {Department of Physics, Indira Gandhi Institute of Technology, Sarang, Dhenkanal, Odisha-759146, India, Email: tripathy\_sunil@rediffmail.com}}
	}

\affiliation{ }
                
%%%%%%%%%%%%

\begin{abstract}

The energy and momentum of the   Bianchi type $III$  universes are obtained using different prescriptions for the energy-momentum complexes in the framework of General Relativity.  The energy and momentum of the Bianchi $III$ universe is found to be zero for  the M\o{}ller prescription. For all other prescriptions the energy and momentum vanish when  the metric parameter $h$ vanishes. In an earlier work, Tripathy et al. \cite{SKT15} have obtained the energy and momentum of Bianchi $VI_h$ metric and found that the energy of the Universe vanish only for $h=-1$. This result raised a question: why this specific choice?. We explored the Tryon's conjecture that 'the Universe must have a zero net value for all conserved quantities' to get some ideas on the specific values of this parameter for Bianchi type Universes.

\end{abstract}


%%%%%


\maketitle

%%%%%%%%%%%%%%%%%%%%%%%%%%%%%%%%%%%%%%%%%%

\section{Introduction}

The General Relativity(GR) as formulated by Einstein is now hundred years old, but the problem of energy-momentum localization in GR has not yet been settled. Einstein conceived the idea of covariant conservation of energy and momenta of gravitational fields along with those of matter and non gravitational fields\cite{Ein15}. However, quantities like energy and momentum at any local point of a manifold should always be conserved as per the usual conservation law where an ordinary differentiation of the energy momentum tensor $T^{j}_i$ should vanish i.e. $T_{i,j}^j=0$. The covariant formulation requires nontensorial fields. Obviously, the energy-momentum due to the gravitational field turns out to be non tensorial (pseudo tensor). The choice of this pseudo tensor is not unique and therefore it has led to the formulation of a number of prescriptions for the calculation of energy and momentum \cite{Tol34, Pap48, LL51, Berg53, Wein72, Moll58, Komar, Pen82}. The interesting thing about these prescriptions is that they depend on the coordinate systems used. It has been observed earlier that, for quasi-Cartesian coordinates, all the prescriptions can provide some reasonable and meaningful results. However some coordinate independent energy-momentum complexes have been proposed by M\o{}ller \cite{Moll58}, Komar \cite{Komar} and Penrose \cite{Pen82}. But some of these coordinate dependent prescriptions are questioned for their limited applicability.

The issue of energy localization has been widely discussed in literature in the frame work of both General Relativity and teleparallel gravity. Misner et al. showed that energy is localized only  for spherically symmetric systems\cite{Misner73}. Cooperstock and Sarracino counter commented the idea of Misner and established that if energy is localized in spherically symmetric system then it can be localizable in any space time \cite{Coop78}. Bondi perceived that, a non localizable form of energy is not admissible in General Relativity, because any form of energy contributes to gravitation and therefore its location can in principle be found \cite{Bondi90}. Virbhadra and his collaborators revived the debate and proved that energy-momentum complexes coincide and give reasonable results for some well-known and physically significant space-times \cite{Vir99, Rose, KSV}. Virbhadra showed that different prescriptions can provide same results when Kerr-Schild Cartesian coordinates are used \cite{Vir99}. Followed by Virbhadra, many researchers obtained interesting results on this pressing issue of energy localisation \cite{others}. Contrary to the previous results, Gad explored about the failure of the agreement in some specific examples of space times \cite{Gad04, Gad04a, Gad04b, Gad06}. Amidst the failures to settle the issue in the context of General Relativity, the energy-momentum has also been formulated in the frame work of teleparallel gravity and has attracted a lot of attention in recent times \cite{Vargas04, Salti05, Ayd05, Sharif10, Gad07, Aygun15}. It has been concluded in some recent works that, the energy-momentum definitions are identical not only in General Relativity but also in teleparallel gravity \cite{Gad07,Aygun07, Aygun08}. 

Tryon anticipated  the net energy of the Universe to be zero\cite{Tryon}.  Albrow \cite{Albrow73} also had a similar assumption on the net energy of the Universe. Rosen from a calculation of the net energy of a closed homogeneous isotropic universe described by a Friedmann-Robertson-Walker (FRW) metric using Einstein energy-momentum complex showed that the total energy of the Universe is zero everywhere \cite{Rosen94}.  Cooperstock and Israelit also found similar results for closed FRW Universe \cite{Coop95}. Johri et al.  also obtained similar results for a closed FRW Universe in Landau-Lifshitz complex \cite{Johri95}. Vargas calculated the energy-momentum of FRW Universe in Landau-Lifshitz and Einstein prescription in the context of teleparallel gravity and obtained zero total energy of the Universe. In a recent work, Tripathy et al.\cite{SKT15} have obtained the energy and momentum of Bianchi type $VI_h$ ($BVI_h$) Universes in the framework of General Relativity in different prescriptions and have shown that the results can only agree for a specific value of the metric parameter $h$. They have also questioned on the basis of Tryon's conjecture that, why the specific spacetime requires the specific value of the parameter. In the present work, we have tried to investigate upon the question raised by Tripathy et al. by considering  anisotropic Bianchi type Universes. It is worth to mention here that, anisotropic spacetimes are more interesting to investigate in the context of recent observations. Many authors have taken interest in the calculation of the energy and momentum of anisotropic Universe in recent times. Banerjee and Sen \cite{Ban97}, Xulu \cite{Xulu00} and Ayd $\ddot{o}$gdu et al. \cite{Ayd05} have obtained the total energy density of Bianchi type-I Universes to be zero everywhere using different nergy-momentum complexes either in General Relativity or in teleparallel gravity . Also, they have calculated the energy of LRS Bianchi $II$ Universe to have consistent results \cite{AS06}. Radinschi calculated the energy distribution of a Bianchi type $VI_0$ Universe using different prescriptions like Tolman, Bergman and Thomson and M\o{}ller  to find zero total energy of the Universe \cite{Rad00}. In another work, Radinschi calculated the energy of Bianchi type $VI_0$ Universe using the Landu-Lifshitz, Papapetrou and Weinberg prescriptions  and found zero net energy due to matter and fields\cite{Rad01}. Aygun and Tarhan have obtained the energy and momentum of Bianchi $IV$ Universe in different complexes in the framework of both the General Relativity and teleparallel gravity \cite{Aygun12}.

The organisation of the paper is as follows: In Section 2, we present the basics of an anisotropic Bianchi type $III$ Universe. In Section 3, the energy and momentum densities for the anisotropic Bianchi $III$ Universe are obtained using some popular prescriptions. Results of the present work are discussed and analysed basing upon the Tryon's conjecture advocating a null total energy state of the Universe in Section 4. At the end, in Section 5, the summary and conclusion the present work are presented. In the present work, we have used the convention that, the Latin indices take values from $0$ to $3$ and Greek indices run from $1$ to $3$. Also, we have used the geometrized unit system where $8\pi G=c=1$, $G$ and $c$ being the Newtonian Gravitational constant and speed of light in vacuum respectively.

%*****************************************************************************************************************

\section{Anisotropic Bianchi type $III$ Universe}

The Universe is observed to be mostly isotropic and can be well explained by the usual $\Lambda$CDM ( $\Lambda$ dominated Cold Dark Matter) model. However, certain measurements of cosmic microwave background from Wilkinson Microwave Anisotropic Probe (WMAP) show some anomalous features of  $\Lambda$CDM model at large scale \cite{Hinshaw09}. These precise measurements suggest an asymmetric expansion of the Universe with one direction expanding in different manner than the other two transverse directions \cite{SKT14, Buiny06, Watan09}. The Planck data \cite{Ade} shows a slight red shifting of the primordial power spectrum of curvature perturbation from exact scale invariance. It can be inferred from the Planck data that usual $\Lambda$CDM model can not be a good fit at least at high multipoles. The issue of global anisotropy can be dealt in many ways. However, a simple way is to modify the FRW model by considering asymmetric expansion along different spatial directions. In these sense, Bianchi type models play important roles. The Bianchi type models are homogeneous having anisotropic spatial sections and are the  exact solutions of Einstein field equations. In the present work, we have considered the Bianchi type III (BIII) model in its generalised form 

\begin{equation}
ds^{2}=-dt^{2}+A^2(t)dx^{2}+B^2(t)e^{2hx}dy^{2}+C^2(t)dz^{2},\label{eq:1}
\end{equation}
%****************
where $A$, $B$ and $C$ are the directional scale factors considered as functions of cosmic time $t$ only. The present model is considered in such a manner that the exponent $h$ is a constant of time and can assume any real values compatible to the real universe. 

Different non vanishing components of the Einstein tensor $G_{ij}=R_{ij}-\frac{1}{2}Rg_{ij}$ for the above metric are 

\begin{eqnarray}\label{eq:2}
G_{00} &=& -\left[\frac{\dot{A}\dot{B}}{AB}+\frac{\dot{B}\dot{C}}{BC}+\frac{\dot{A}\dot{C}}{AC}+\frac{h^2}{A^2}\right],\\\nonumber
G_{11} &=& A^2\left[\frac{\ddot{B}}{B}+\frac{\ddot{C}}{C}+\frac{\dot{B}\dot{C}}{BC}\right],\\\nonumber
G_{22} &=& B^2e^{2hx}\left[\frac{\ddot{A}}{A}+\frac{\ddot{C}}{C}+\frac{\dot{A}\dot{C}}{AC}\right],\\\nonumber
G_{33} &=& C^2\left[\frac{\ddot{A}}{A}+\frac{\ddot{B}}{B}+\frac{\dot{A}\dot{B}}{AB}-\frac{h^2}{A^2}\right],\\\nonumber
G_{10} &=& h\left(\frac{\dot{B}}{B}-\frac{\dot{A}}{A}\right).
\end{eqnarray}
%****************************************
Here $R_{ij}$ is the Ricci tensor, $R$ is the Ricci scalar, $g_{ij}$ is the metric tensor. Also, $G_{10}=G_{01}= R_{10}$. In the above equations  an overhead dot over a field variable represents  a time derivative.

%%%%%%%%%%%%%%%%%%%%%%%%%%%%%%%%%%%%%%%%%%%%%%%%%%%%%%%%%%%%%%%%%%%%%%%%%%%%%%%%%%%%%%%%%%%%%%%%%%%%%%%%%%%%%%%%%%%
% % % % % % % % % % % % % % % % % % % % % % % % % % %

\section{Energy-Momentum Complexes}

In this section, we present the general results of the energy and momentum of generalised $BIII$ Universe for different energy-momentum complexes namely Einstein, Landau-Lifshitz, Papapetrou, Bergmann-Thompson and M\o{}ller prescriptions. The definition of the energy-momentum pseudo tensors and corresponding formulations for energy and momentum  in different prescriptions are given in Table-I. The results for energy and momentum densities will be presented in general form of the directional scale factors $A, B$, $C$ and the exponent $h$. From these general results, one can easily obtain the energy and momentum by considering the time dependence of the scale factors. It is worth to mention here that, we restrict the definitions of the well known energy-momentum prescriptions to the frame work of General Relativity. In the following subsections we report the derived non vanishing components of the super potentials and the consequent energy and momentum densities. The derived energy and momentum densities are given in Table-II.

%===================================================================================================

\begin{table}
\caption{Different energy-momentum prescriptions used in the present work.}
\centering
\begin{tabular}{l|c|c|c}
\hline \hline
Prescription		&	Energy-momentum Pseudo-tensor 	& Densities 	 		& Energy-momentum 	\\
					&									&						&	four vector$(P_i)$		   	\\
\hline
Einstein			&	$\Theta_i^k=\frac{1}{16\pi}H_{i,l}^{kl}$	&	$\Theta_0^0$, 	& 	$\int\int\int \Theta_i^0 dx^1dx^2dx^3,$  \\
					&	$H_i^{kl}=-H_i^{lk}=\frac{g_{in}}{\sqrt{-g}}[-g(g^{kn}g^{lm}-g^{ln}g^{km})]$	&	$\Theta_\alpha^0$	& 	\\
\hline
Landau-Lifshitz		&$L^{ik}=\frac{1}{16\pi}\lambda_{,lm}^{iklm}$	&	$L^{00}$,	& $\int\int\int L^{i0} dx^1dx^2dx^3$,	\\
					&	$\lambda^{iklm}=-g(g^{ik}g^{lm}-g^{il}g^{km})$	&	$L^{\alpha 0}$	& 	\\\hline
Papapetrou			&	$\Sigma^{ik}=\frac{1}{16\pi} {\mathcal{N}}^{iklm}_{,lm}$	&	$\Sigma^{00}$,	& $\int\int\int \Sigma^{i0} dx^1dx^2dx^3$,\\
					&	${\mathcal{N}}^{iklm}=\sqrt{-g}$		&	$\Sigma^{\alpha 0}$	& \\
					&$\times (g^{ik}\eta^{lm}-g^{il}\eta^{km}+g^{lm}\eta^{ik}-g^{lk}\eta^{im})$&&\\\hline
Bergmann-Thompson	&${\bf{B}}^{ik}=\frac{1}{16\pi} [g^{il}{\mathcal{B}}^{km}_l]_{,m}$	&	${\bf{B}}^{00}$,	& $\int\int\int {\bf{B}}^{i0} dx^1dx^2dx^3$,\\
					&	${\mathcal{B}}_l^{km}=\frac{g_{ln}}{\sqrt{-g}} [-g(g^{kn}g^{mp}-g^{mn}g^{kp})]_{,p}$	&	${\bf{B}}^{\alpha 0}$	& \\\hline
M\o{}ller			&	$T_i^k = \frac{1}{8\pi}\chi_{i,l}^{kl}$	&	$T_0^0$,	& $\int\int\int T_i^0 dx^1dx^2dx^3$	\\
					&	${\chi_{i}^{kl}} = -\chi_{i}^{lk} = \sqrt{-g}[g_{in,m}-g_{im,n}]g^{km}g^{nl}$	&	$T_{\alpha}^0$	& \\
\hline
\end{tabular}
\end{table}

%=========================================================================================================

\subsection{Einstein Energy-Momentum Complex}

The required non-vanishing components of the $ H_i^{kl}$ are 
%***************************
\begin{eqnarray}
H_0^{01} &=& \frac{2BCh}{A}e^{hx},\nonumber\\
H_1^{01} &=& 2ABC \left(\frac{\dot{B}}{B}+\frac{\dot{C}}{C}\right)e^{hx},\nonumber\\
H_2^{02} &=& 2ABC \left(\frac{\dot{A}}{A}+\frac{\dot{C}}{C}\right)e^{hx},\nonumber\\
H_3^{03} &=& 2ABC \left(\frac{\dot{A}}{A}+\frac{\dot{B}}{B}\right)e^{hx},\nonumber\\
H_3^{31} &=& \frac{2BCh}{A}e^{hx}.\label{eq:3} 
\end{eqnarray}
%********************************
The components of energy and momentum densities can now be obtained as
%*********************************
\begin{eqnarray}
\Theta_0^0 &=& \frac{h^2}{8\pi}\frac{BC}{A}e^{hx},\nonumber\\
\Theta_1^0 &=& \frac{h}{8\pi}ABC\left(\frac{\dot{B}}{B}+\frac{\dot{C}}{C}\right) e^{hx},\nonumber\\
\Theta_2^0 &=& \Theta_3^0=0.\label{eq:4}
\end{eqnarray}
%*************************************

\subsection{Landau and Lifshitz Energy-Momentum Complex}

For the the generalised $BIII$ model, the non-vanishing components of $\lambda^{iklm}$ are obtained as 

\begin{eqnarray}
\lambda^{0011} &=&\lambda^{1100} = -B^2C^2e^{2hx},\nonumber\\
\lambda^{0022} &=&\lambda^{2200} = -A^2C^2,\nonumber\\
\lambda^{1010} &=&\lambda^{0101} = B^2C^2e^{2hx},\nonumber\\
\lambda^{1122} &=&\lambda^{2211} = C^2,\nonumber\\
\lambda^{2233} &=&\lambda^{3322} = A^2,\nonumber\\
\lambda^{2323} &=&\lambda^{3232} = -A^2,\nonumber\\
\lambda^{1133} &=&\lambda^{3311} = B^2e^{2hx},\nonumber\\
\lambda^{1212} &=&\lambda^{2121} = -C^2,\nonumber\\
\lambda^{1313} &=&\lambda^{3131} = -B^2e^{2hx},\nonumber\\
\lambda^{2020} &=&\lambda^{0202} = A^2C^2,\nonumber\\
\lambda^{0033} &=&\lambda^{3300} = -A^2B^2e^{2hx},\nonumber\\
\lambda^{3030} &=&\lambda^{0303} = A^2B^2e^{2hx}.\label{eq:5}
\end{eqnarray}
%********************************

Consequently, the energy and momentum densities in the Landau and Lifshitz prescription become

\begin{eqnarray}
L^{00} &=& -\frac{h^2}{4\pi}B^2C^2e^{2hx},\nonumber\\
L^{10} &=& \frac{h}{4\pi}BC(\dot{B}C+B\dot{C})e^{2hx},\nonumber\\
L^{20} &=& L^{30} =0.\label{eq:6}
\end{eqnarray}
%***************************************
\subsection{Papapetrou Energy-Momentum Complex}

The  required non-vanishing components of ${\mathcal{N}}^{iklm}$ for the calculation of the energy ($\Sigma^{00}$) and momentum density ($\Sigma^{\alpha 0}$) components are
%************************************
\begin{eqnarray}
{\mathcal{N}}^{0101} &=& {\mathcal{N}}^{1001} = ABCe^{hx},\nonumber\\
{\mathcal{N}}^{0011} &=& {\mathcal{N}}^{1100} = -\left(1+\frac{1}{A^2}\right)ABCe^{hx},\nonumber\\
{\mathcal{N}}^{0110} &=& {\mathcal{N}}^{1010} = \left(\frac{BC}{A}\right)e^{hx},\nonumber\\
{\mathcal{N}}^{0022} &=& {\mathcal{N}}^{2200} = -\left(1+\frac{1}{B^2e^{2hx}}\right)ABCe^{hx},\nonumber\\
{\mathcal{N}}^{0330} &=& {\mathcal{N}}^{3030} = \frac{AB}{C}e^{hx},\nonumber\\
{\mathcal{N}}^{0220} &=& {\mathcal{N}}^{2020} = \left(\frac{AC}{B}\right)\frac{1}{e^{hx}},\nonumber\\
{\mathcal{N}}^{1122} &=& {\mathcal{N}}^{2211} = \left(\frac{1}{B^2e^{2hx}}+\frac{1}{A^2}\right)ABCe^{hx},\nonumber\\
{\mathcal{N}}^{1221} &=& {\mathcal{N}}^{2121} = -\left(\frac{AC}{B}\right)\frac{1}{e^{hx}},\nonumber\\
{\mathcal{N}}^{1331} &=& {\mathcal{N}}^{3131} = -\left(\frac{AB}{C}\right)e^{hx},\nonumber\\
{\mathcal{N}}^{1313} &=& {\mathcal{N}}^{3113} = -\left(\frac{BC}{A}\right)e^{hx},\nonumber\\
{\mathcal{N}}^{1133} &=& {\mathcal{N}}^{3311} = \left(\frac{1}{A^2}+\frac{1}{C^2}\right)ABCe^{hx},\nonumber\\
{\mathcal{N}}^{2002} &=& {\mathcal{N}}^{3003} = ABCe^{hx},\nonumber\\
{\mathcal{N}}^{2112} &=& -{\mathcal{N}}^{1212} = \left(\frac{BC}{A}\right)e^{hx},\nonumber\\
{\mathcal{N}}^{2332} &=& {\mathcal{N}}^{3232} = -\left(\frac{AB}{C}\right)e^{hx},\nonumber\\
{\mathcal{N}}^{3300} &=& {\mathcal{N}}^{0033} = -\left(1+\frac{1}{C^2}\right)ABCe^{hx},\nonumber\\
{\mathcal{N}}^{3322} &=& {\mathcal{N}}^{2233} = \left(\frac{1}{c^2}+\frac{1}{B^2e^{2hx}}\right)ABCe^{hx}.\label{eq:7}
\end{eqnarray}
%**********************************
The energy density and momentum density components in the Papapetrou prescription are obtained from  \eqref{eq:7} as 

\begin{eqnarray}
\Sigma^{00} &=& -\frac{h^2}{16\pi}\left(1+\frac{1}{A^2}\right)ABC e^{hx},\nonumber\\
\Sigma^{10} &=& \frac{h}{16\pi}\left[\left(\frac{\dot{A}}{A}+\frac{\dot{B}}{B}+\frac{\dot{C}}{C}\right)+\left(\frac{1}{A^2}\right)\left(\frac{\dot{B}}{B}+\frac{\dot{C}}{C}-\frac{\dot{A}}{A}\right)\right] ABCe^{hx},\nonumber\\
\Sigma^{20} &=& 0, \nonumber\\
\Sigma^{30} &=& 0. \label{eq:8}
\end{eqnarray}

%***************************************************
\subsection{Bergmann-Thompson Energy-Momentum Complex}
The non-vanishing components of $\mathcal{B}^{km}_l$ for $BIII$ Universe are
%******************************************
\begin{eqnarray}
{\mathcal{B}}_0^{01} &=& -{\mathcal{B}}_0^{10} = {\mathcal{B}}_3^{31} = -{\mathcal{B}}_3^{13} = \frac{2BCh}{A}e^{hx},\nonumber\\
{\mathcal{B}}_1^{01} &=& -{\mathcal{B}}_1^{10} = 2A(\dot{B}C+B\dot{C})e^{hx},\nonumber\\
{\mathcal{B}}_2^{02} &=& -{\mathcal{B}}_2^{20} = 2B(\dot{A}C+A\dot{C})e^{hx},\nonumber\\ 
{\mathcal{B}}_3^{03} &=& -{\mathcal{B}}_3^{03} = 2C(\dot{A}B+A\dot{B})e^{hx}.\label{eq:9}
\end{eqnarray}
%*******************************************

Using eq. \eqref{eq:9}, the energy and momentum density components i.e. ${\bf{B}}^{00}$ and ${\bf{B}}^{\alpha 0}$, can be obtained as,

\begin{eqnarray}
{\bf{B}}^{00} &=& \frac{h^2}{8\pi}\frac{BC}{A}e^{hx},\nonumber\\
{\bf{B}}^{10} &=& \frac{h}{8\pi}\frac{BC}{A}\left(\frac{\dot{B}}{B}+\frac{\dot{C}}{C}\right)e^{hx}\nonumber\\
{\bf{B}}^{20} &=& {\bf{B}}^{30} = 0\label{eq:10}
\end{eqnarray}
%******************************************
\subsection{M\o{}ller Energy Momentum Complex}

The non-vanishing components of ${\chi_{i}^{kl}}$ are
%******************************************
\begin{eqnarray}
{\chi_{1}^{01}} &=& -{\chi_{1}^{10}} = -2\dot{A}BCe^{hx},\nonumber\\
{\chi_{2}^{02}} &=& -{\chi_{2}^{20}} = -2A\dot{B}Ce^{hx},\nonumber\\
{\chi_{3}^{03}} &=& -{\chi_{3}^{30}} = -2AB\dot{C}e^{hx},\nonumber\\
{\chi_{2}^{21}} &=& -{\chi_{2}^{12}} = -\frac{2BC}{A}e^{hx},\nonumber\\
{\chi_{3}^{31}} &=& -{\chi_{3}^{13}} = -\frac{2BCh}{A}e^{hx}.\label{eq.11}
\end{eqnarray}
%*******************************************
The energy and momentum density components for M\o{}ller energy-momentum complex are obtained as,

\begin{eqnarray}
T_0^0 &=& T_2^0 = T_3^0 =0,\nonumber\\
T_1^0 &=& -\frac{1}{4\pi}\dot{A}BChe^{hx}\label{eq.12}
\end{eqnarray}
%*******************************************
%===================================================================================================

\begin{table}
\caption{Derived expressions for the energy and momentum densities for different energy-momentum prescription used in the present work.}
\centering
\begin{tabular}{l|l|l}
\hline \hline
Prescription		&	Energy-momentum Pseudo-tensor 	& Densities 	 		\\
					&									&							   	\\
\hline
Einstein			&	$\Theta_i^k=\frac{1}{16\pi}H_{i,l}^{kl}$	&	$\Theta_0^0 = \frac{h^2}{8\pi}\frac{BC}{A}e^{hx}$,\\
					&$H_i^{kl}=-H_i^{lk}=\frac{g_{in}}{\sqrt{-g}}[-g(g^{kn}g^{lm}-g^{ln}g^{km})]$		&$\Theta_1^0 = \frac{h}{8\pi}ABC\left(\frac{\dot{B}}{B}+\frac{\dot{C}}{C}\right) e^{hx}$, 	 \\
					&		&	$\Theta_2^0 =\Theta_3^0=0$ 		\\
\hline
Landau-Lifshitz		& $L^{ik}=\frac{1}{16\pi}\lambda_{,lm}^{iklm}$	&	 $L^{00} = -\frac{h^2}{4\pi}B^2C^2e^{2hx}$,		\\
					&$\lambda^{iklm}=-g(g^{ik}g^{lm}-g^{il}g^{km})$	&	$L^{10} = \frac{h}{4\pi}BC(\dot{B}C+B\dot{C})e^{2hx}$,\\
					&		&	$L^{20} = L^{30} =0$	\\\hline
Papapetrou			&	$\Sigma^{ik}=\frac{1}{16\pi} {\mathcal{N}}^{iklm}_{,lm}$	&	$\Sigma^{00}= -\frac{h^2}{16\pi}\left(1+\frac{1}{A^2}\right)ABC e^{hx}$,\\
					&	&$\Sigma^{10} = \frac{h}{16\pi}\left(\frac{\dot{A}}{A}+\frac{\dot{B}}{B}+\frac{\dot{C}}{C}\right)ABCe^{hx}$\\
					&	${\mathcal{N}}^{iklm}=\sqrt{-g}$	&$+\frac{h}{16\pi}\left(\frac{1}{A^2}\right)\left(\frac{\dot{B}}{B}+\frac{\dot{C}}{C}-\frac{\dot{A}}{A}\right) ABCe^{hx}$,	 \\
					&$\times (g^{ik}\eta^{lm}-g^{il}\eta^{km}+g^{lm}\eta^{ik}-g^{lk}\eta^{im})$	&$\Sigma^{20} =\Sigma^{30} =0$\\\hline
Bergmann-Thompson	&${\bf{B}}^{ik}=\frac{1}{16\pi} [g^{il}{\mathcal{B}}^{km}_l]_{,m}$	&	${\bf{B}}^{00}=\frac{h^2}{8\pi}\frac{BC}{A}e^{hx}$,\\
					&	${\mathcal{B}}_l^{km}=\frac{g_{ln}}{\sqrt{-g}} [-g(g^{kn}g^{mp}-g^{mn}g^{kp})]_{,p}$	&	${\bf{B}}^{10}=\frac{h}{8\pi}\frac{BC}{A}\left(\frac{\dot{B}}{B}+\frac{\dot{C}}{C}\right)e^{hx}$,	 \\
					&&${\bf{B}}^{20} ={\bf{B}}^{30} = 0$\\\hline
M\o{}ller			&	$T_i^k = \frac{1}{8\pi}\chi_{i,l}^{kl}$	&	$T_0^0 = T_2^0 = T_3^0 =0$,		\\
					&	${\chi_{i}^{kl}} = -\chi_{i}^{lk} = \sqrt{-g}[g_{in,m}-g_{im,n}]g^{km}g^{nl}$	&	$T_1^0 = -\frac{1}{4\pi}\dot{A}BChe^{hx}$	 \\
\hline
\end{tabular}
\end{table}

%=========================================================================================================
\section{Results and Discussion}
From the results of the energy and momentum densities for $BIII$ Universe in different well known prescriptions, one can note that, the energy and momentum densities in M\o{}ller prescription is zero and is independent of the value of the exponent $h$. However, the momentum component in M\o{}ller prescription vanishes only for $h=0$. In all other cases, the energy and momentum do not vanish identically and depend on the exponent $h$. For $h=0$, the energy and momentum in these complexes vanish. It is interesting to note that, the energy and momentum densities for the $BIII$ Universe are the same for the Einstein and Bergmann-Thompson prescriptions. Also, the energy for the said model is negative in the Landau-Lifshitz and the Papapetrou prescriptions for non zero values of the exponent $h$.

In a recent work, Tripathy et al. \cite{SKT15} have calculated the energy and momentum of Bianchi $VI_h$ Universes ($BVI_h$) in Einstein, Landau-Lifshitz, Papapetrou and Bergmann-Thompson prescriptions. In that work the authors have observed that, the energy and momentum of the Universe vanish for a specific choice of the exponent $h=-1$. In order to complete the work of Tripathy et al. and to derive a conclusive remark on the exponent $h$, based upon Tryon's conjecture, we have calculated the energy of Bianchi $VI_h$ Universe in the M\o{}ller prescription.

The Bianchi $VI_h$ Universes are modelled through the metric 

\begin{equation}
ds^{2}=-dt^{2}+A^2(t)dx^{2}+B^2(t)e^{2x}dy^{2}+C^2(t)e^{2hx}dz^{2}.\label{eq:13}
\end{equation}

The non vanishing components of ${\chi_{i}^{kl}}$ for Bianchi $VI_h$ Universes  are
%******************************************
\begin{eqnarray}
{\chi_{1}^{01}} &=& -{\chi_{1}^{10}} = -2\dot{A}BCe^{(1+h)x},\nonumber\\
{\chi_{2}^{02}} &=& -{\chi_{2}^{20}} = -2A\dot{B}Ce^{(1+h)x},\nonumber\\
{\chi_{3}^{03}} &=& -{\chi_{3}^{30}} = -2AB\dot{C}e^{(1+h)x},\nonumber\\
{\chi_{2}^{21}} &=& -{\chi_{2}^{12}} = -\frac{2BC}{A}e^{(1+h)x}\nonumber\\
{\chi_{3}^{31}} &=& -{\chi_{3}^{13}} = -\frac{2BCh}{A}e^{(1+h)x}.\label{eq.14}
\end{eqnarray}

Consequently, the energy and momentum density components for M\o{}ller energy-momentum complex are 

\begin{eqnarray}
T_0^0 &=& T_2^0 = T_3^0 =0,\nonumber\\
T_1^0 &=& -\frac{1+h}{4\pi}\dot{A}BCe^{hx}.\label{eq.15}
\end{eqnarray}

The M\o{}ller energy for Bianchi $VI_h$ Universe is zero. However, the momentum of this model vanish only for $h=-1$.

In his interesting work, Tryon \cite{Tryon} assumed that the Universe has appeared from nowhere about $10^{10}$ years ago. As per his thought, at the time of creation of the Universe, the conventional laws of Physics may not have been violated. Tryon has proposed a Big Bang model where the Universe was emerged from a large scale quantum fluctuation of the vacuum.  His model predicted a Universe that is homogeneous, isotropic and closed consisting of equal amount of matter and anti-matter. It is worth to mention here that, the Big Bang model and the predictions are consistent with the observations from Cosmic Microwave Background Radiations (CMB). In order to emphasize his thought, Tryon proposed a remarkable conjecture on the energy-momentum of the Universe according which the Universe must have a zero net value for all conserved quantities. In the same paper, Tryon  has mentioned that any closed Universe has zero energy. He substantiated his idea by arguments. Many authors have claimed that the laws of Physics could have created the Universe from nothing \cite{Albrow73, Krauss12, Hawking96, Hebert12}.  Xulu \cite{Xulu2000} studied energy and momentum in Bianchi type I Universes and his results supported the conjecture of Tryon. Berman, from different arguments, has also shown that the Robertson-Walker's Universe and any other Machian ones have zero total energy \cite{Berman07, Berman07a, Berman09}.

\begin{table}
\caption{Energy and momentum densities of diagonal and anisotropic generalised Bianchi type Universes using different energy-momentum prescriptions used in the present work.}
\centering
\begin{tabular}{l|c|l}
\hline \hline
Prescription		&	 Energy  Density 	 & Momentum Densities		\\
					&						 &									   	\\
\hline
Einstein				&	$\Theta_0^0 = \frac{(\alpha+\beta)^2}{8\pi}\frac{BC}{A}e^{(\alpha+\beta)x}$ &$\Theta_1^0 = \frac{(\alpha+\beta)}{8\pi}ABC\left(\frac{\dot{B}}{B}+\frac{\dot{C}}{C}\right) e^{(\alpha+\beta)x},$\\
				& 	&	$\Theta_2^0 =\Theta_3^0=0$ 		\\
\hline
Landau-Lifshitz		&	 $L^{00} = -\frac{(\alpha+\beta)^2}{4\pi}B^2C^2e^{2(\alpha+\beta)x}$&	$L^{10} = \frac{(\alpha+\beta)}{4\pi}BC(\dot{B}C+B\dot{C})e^{2(\alpha+\beta)x}$,\\
					&&	$L^{20} = L^{30} =0$		\\\hline
Papapetrou			&	$\Sigma^{00}= -\frac{(\alpha+\beta)^2}{16\pi}\left(1+\frac{1}{A^2}\right)ABC e^{(\alpha+\beta)x}$&$\Sigma^{10} = \frac{(\alpha+\beta)}{16\pi}\left(\frac{\dot{A}}{A}+\frac{\dot{B}}{B}+\frac{\dot{C}}{C}\right)ABCe^{(\alpha+\beta)x}$\\
					&
					&$+\frac{(\alpha+\beta)}{16\pi}\left(\frac{1}{A^2}\right)\left(\frac{\dot{B}}{B}+\frac{\dot{C}}{C}-\frac{\dot{A}}{A}\right) ABCe^{(\alpha+\beta)x}$,	\\
					&&$\Sigma^{20} =\Sigma^{30} =0$\\\hline
Bergmann-Thompson	&	${\bf{B}}^{00}=\frac{(\alpha+\beta)^2}{8\pi}\frac{BC}{A}e^{(\alpha+\beta)x}$&${\bf{B}}^{10}=\frac{(\alpha+\beta)}{8\pi}\frac{BC}{A}\left(\frac{\dot{B}}{B}+\frac{\dot{C}}{C}\right)e^{(\alpha+\beta)x}$\\
					&&${\bf{B}}^{20} ={\bf{B}}^{30} = 0$\\\hline
M\o{}ller			&	$T_0^0 = 0$		&$T_1^0 = -\frac{(\alpha+\beta)}{4\pi}\dot{A}BCe^{(\alpha+\beta)x}$\\
					&		 &$T_2^0 = T_3^0 =0$\\
\hline
\end{tabular}
\end{table}

More or less, it is now an accepted fact that, our Universe is created out of nothing and its net energy is zero. If this conjecture is to be valid then all prescriptions for energy-momentum should agree with null values of the energy and momentum densities. In view of this and the calculations of energy and momentum for different Bianchi type Universes, it is certain that, a flat model is necessary. In order to substantiate our view, we present the energy and momentum densities for a more general Bianchi type Universe represented through the metric

\begin{equation}
ds^{2}=-dt^{2}+A^2(t)dx^{2}+B^2(t)e^{\alpha x}dy^{2}+C^2(t)e^{\beta x}dz^{2},\label{eq:16}
\end{equation}
in Table-III. For Bianchi type Universes represented by the above metric, vanishing of the total energy of the Universe requires that, the sum $\alpha+\beta$ should vanish i.e $\alpha+\beta=0$. According to this rule of vanishing total exponents (multiplied to $x$), in Bianchi III Universe with $\alpha=h$ and $\beta=0$, we require that, the exponent $h$ should vanish. Similarly, the question raised in the work of Tripathy et al.\cite{SKT15} that "why $h=-1$ in Bianchi $VI_h$ is so special?" can now be answered. In Bianchi $VI_h$ Universe, in order to make the total exponent to vanish, $h$ has to take a value of $-1$. This is in accordance with the Tryon's conjecture.

\section{Summary}
In the present work, we have obtained the energy and momentum distributions for an anisotropic Bianchi type III Universe in some well known prescriptions such as Einstein, Landau-Liftshitz, Papapetrou, Bergmann-Thompson and M\o{}ller energy-momentum complexes in General Relativity. It is obseved that, M\o{}ller energy vanishes for any value of the exponent $h$ appearing in the $BIII$ metric. However in all other cases, the energy and momentum of the Universe vanish for the specific choice $h=0$. Also, the energy densities for the Einstein prescription and the Bergmann-Thompson prescription are found to be the same. 

In a recent work, Tripathy et al.\cite{SKT15} have calculated the energy and momentum of Bianchi $VI_h$ Universes and have obtained that, for the specific choice of the exponent $h=-1$ appearing in the metric, all prescriptions considered in that work provide null total energy. Based upon their result they have raised a question that:"Why is the case $h=-1$ case so special?". In the present work, we have investigated upon that question. In order to get a conclusive remark and for completeness, we have calculated the M\o{}ller energy and momentum for Bianchi $VI_h$ Universes. We observed that, like the $BIII$ Universe, the M\o{}ller energy for $BVI_h$ Universes vanishes and does not depend on the exponent $h$. But the momentum components depend upon the exponent $h$.

In view of the Tryon's conjecture that advocates a zero total energy of the Universe, it is suggested in the present work that, the Bianchi type Universes described through a general metric in \eqref{eq:16} require that the sum $\alpha+\beta$ should vanish. Therefore, in the present work for $BIII$ Universe, the exponent $h$ requires a value of 0 whereas for $BVI_h$ Universes (as in the work of Tripathy et al.\cite{SKT15}), the exponent $h$ requires a value $-1$. Even if, the zero total energy of the Universe or the creation of the present Universe from nothing has been an accepted fact in the context of Big Bang cosmology, still it remains as open problem and a subject of intense debate. Many other cosmological models else than Big Bang models have been proposed in recent times. In view of this, our results are important in the sense that, they may put some insights into the old and unsettled problem of energy-momentum calculation. It is certain that, under the purview of General Relativity, all the prescriptions should provide the same result for the energy of the Universe and in this sense, anisotropic Bianchi type Universes must also be consistent to that.

%==================================================================================================================
\newpage

%++++++++++++++++++++++++++++++++++++++++++++++++++++++++++++++++++++++++++++++++++++++++++++++++
\section*{Conflict of Interests}
The authors declare that there is no conflict of interests regarding the publication of this paper.

%%%%%%%%%%%%%%%%%%%%%%%%%%%%%%%%%%%%%%%%%%%%%%%%%%%%%%%%%%%%%%%%%%%%%%%%%%%%%%%%%%%%%%%%%%%%%%%%%%%%%%%%%%%%%%%%%%%%%%%%%%%%%
%\newpage

\end{document}